\documentclass[preprint,a4paper,3p,10pt,times,twocolumn,dvipdfm]{elsarticle}
\usepackage{graphics}
\usepackage{graphicx}
\usepackage{epsfig}
\usepackage{amssymb}
\usepackage{amsthm}
\usepackage{natbib}
\usepackage{latexsym}
\usepackage{mathrsfs}
\usepackage{amsmath}
\usepackage{amscd}
\usepackage{color}
\usepackage{verbatim}
\biboptions{sort&compress}
\journal{Physica A}
\begin{document}
\begin{frontmatter}
\title{Characterizing weak chaos in nonintegrable Hamiltonian systems: \\
  the fundamental role of stickiness and initial conditions} 
\author[udesc]{C.~Manchein}
\address[udesc]{Departamento de F\'\i sica, Universidade do Estado 
de Santa Catarina, 89219-710 Joinville, Brazil}
\ead{cesar.manchein@udesc.br}
\author[ufpr,mpipks]{M.~W.~Beims}
\ead{mbeims@fisica.ufpr.br}
\address[ufpr]{Departamento de F\'isica, Universidade Federal do Paran\'a, Caixa Postal 19044, 81531-980, Curitiba, PR, Brazil}
\author[mpipks]{J.~M.~Rost}
\address[mpipks]{Max Planck Institute for the Physics of Complex Systems,
         N\"othnitzer Strasse 38, 01187 Dresden, Germany}
\date{\today}
%
\begin{abstract}
Weak chaos in high-dimensional conservative systems can be characterized 
through sticky effect induced by invariant structures on chaotic 
trajectories. Suitable quantities for this characterization are the 
higher cummulants of the finite time Lyapunov exponents (FTLEs) distribution. 
They gather the {\it whole} phase space relevant dynamics in {\it one} quantity 
and give informations about ordered and random states. This is analyzed here
for discrete Hamiltonian systems with local and global couplings. 
It is also shown that FTLEs plotted {\it versus} initial condition (IC) 
and the nonlinear parameter is essential to understand the 
fundamental role of ICs in the dynamics of weakly chaotic Hamiltonian systems.
\end{abstract}

\end{frontmatter}

\section{Introduction}
\label{Introduction}
The phase space of nonlinear conservative Hamiltonian systems 
can present regions of regular, mixed and chaotic motion,
depending on the nonlinear parameter. The regular region is 
characterized by the complete absence of chaotic trajectories 
while the chaotic region by the absence of regular trajectories,
with exceptions of measure zero stable orbits. The mixed regions 
on the other hand, contains simultaneously the regular and chaotic 
motion (or quasi-regular) and everything becomes complicated 
\cite{zas88,meiss92}: the positions of regular structures, their size 
distribution, exit times, the shapes of the structures boundaries and 
the penetration inside them which is possible in higher dimensions.
Weak chaos occurs in the mixed region where an approximately even 
competition between the regular and chaotic dynamics occurs 
\cite{zas88}. In $2-$dimensional conservative problems ($4-$dimensional 
phase space) the dynamics can be described by using the technique 
of Poincar\'e Surfaces of Section (PSS). However, for high-dimensional 
systems this technique is only partially useful due to the restriction 
of plots to $2-$ and $3-$dimensions, which makes it almost impossible 
to construct adequate PSSs which allow to describe the whole dynamics. 
Beside that, $3-$dimensional plots of high-dimensional systems 
are fairly unsatisfactory projections from the whole system. A recent
work proposes visualization of classical structures using phase space 
slices \cite{martin13}.

Another possibility is to use Lyapunov quantifiers which decide if a 
trajectory is chaotic or not.  In the mixed region stickiness
\cite{zas02,denis12} 
affects the convergence of finite time Lyapunov  exponents (FTLEs),
but they contain essential informations about the properties of the regular 
structures which live in the  high-dimensional phase space. The properties 
of regular structures have been addressed on the same constant energy 
\cite{pettini84}, by studying their dimensionality 
\cite{tsiganis00,malagoli86},  and the almost invariant sets in continuous 
problems \cite{froyland09,dellnitz97}, to mention a few. The purpose of the 
present work is not to describe the invariant structures itself, but to 
quantify their effect on the dynamics. As explained below, it turns out that 
FTLEs are very suitable to do the job in the mixed regions.
Any dynamical evolution of the system depends on the starting point in phase 
space and on the shape and number of regular structures of its surroundings.
Since FTLEs are usually strongly dependent on the initial conditions (ICs) 
and on the sticky motion, they can be used to quantify the amount of regular
motion in high-dimensional phase spaces. It would be nice to check the 
stickiness influence on the smaller alignment index \cite{skokos04,skokos06} 
which rapidly distinguish between ordered and 
chaotic trajectories in Hamiltonian flows.

A very appropriate way to quantify the sticky motion is by analyzing 
higher order cummulants of the FTLEs distribution. This was proposed 
\cite{steven07} for the standard map and applied \cite{cesar-rost11} 
to higher dimensions in conservative non-Hamiltonian systems. It was 
found \cite{cesar-rost11} by this technique that for $2,4,10$ and $20$ 
phase space dimensions, conservative coupled standard maps with 
unidirectional local coupling can be characterized of 
being chaotic, quasi-regular or regular. In addition, for some values of the 
nonlinear parameter in the quasi-regular region, stickiness is shown to affect 
{\it all} unstable directions  {\it simultaneously} and by the {\it same} 
amount, which is quantified by the cummulants mentioned above. This 
remarkable property was named {\it common} behavior \cite{cesar-rost11} 
and its main property is that regular structures in phase space ``attract'' 
the chaotic trajectories by the amount in all unstable directions. Once the
chaotic trajectory is attracted, it remains sticked to the regular structure
and then presents the clustering behavior observed  \cite{woellner09} 
in conservative maps.
But the essential new property from the common behavior 
is that the rate of attraction of the chaotic trajectory into the regular
structure is equal in different unstable directions. 

The question remains if the common motion is a general property found
also in Hamiltonian and symplectic systems, or maybe it depends on the 
particular choice of the coupling? In order to answer this question and
to understand better mixed phase spaces, in this work we extend our 
method  to Hamiltonian systems with global and local couplings and
compare the results. The existence of an additional constant of motion, 
besides the total energy, allows for a clear interpretation of results.

The paper is divided in the following way. In Section \ref{model} the 
local and global coupled maps models used in this work are presented
and in Section \ref{method} we summary the main properties of the
FTLEs distribution in order to detect stickiness. In Section 
\ref{results} we present and discuss the results which are summarized 
in the conclusions in Section \ref{conclusion}.

\section{Lattices of Coupled Hamiltonian Maps}
\label{model}

The model we investigate is conservative with an additional constant 
of motion. It describes $N$ particles coupled on a unit circle, 
where the state of each particle is defined by its position 
$x^{(i)}$ and its conjugate momentum  $p^{(i)}$. The lattice of 
$N$ coupled maps is written as 

\begin{eqnarray}
    \left\{
    \begin{array}{ll}
        p_{t+1}^{(i)} = p_t^{(i)} + f(x_t) \qquad \mbox{mod}\, 1, \\
                      \\
	x_{t+1}^{(i)} = x_t^{(i)} + p_{t+1}^{(i)} \,\,\qquad \mbox{mod}\, 1,
    \end{array}
    \right.
\end{eqnarray}
where $f(x_t)$ can be local or global coupling as discussed next.
The local coupling (LC) is defined accordingly to 

\begin{eqnarray}
 f(x_t) = \frac{K}{2\pi}&&\left\{\sin[2\pi(x_t^{(i+1)}-x_t^{(i)})]\right.\cr
            & & \left. \right.\\
        -&& \left.\sin[2\pi(x_t^{(i)}-x_t^{(i-1)})]\right\}, 
\label{gc}\nonumber
\end{eqnarray}
where $i=1,\ldots,N$. We considered periodic boundary conditions
$p^{(N+1)}=p^{(1)}, x^{(N+1)}=x^{(1)}$.  Some properties of this model
were already investigated numerically in
\cite{kk89,kk90, marcelo12-1}. For the global coupling (GC) we have

\begin{eqnarray}
  f(x_t) =
  \frac{K}{2\pi\sqrt{N-1}}\sum_{j=1,j\not=i}^{N}
\sin[2\pi(x_t^{(j)})-2\pi(x_t^{(i)})], 
\label{lc}
\end{eqnarray}
where $i=1,\ldots,N$ \cite{kk90} . $K$ is simultaneously the nonlinear 
parameter and the coupling strength between distinct sites. For
$K>0$ the interaction 
term $f(x_t)$ between two particles $i$ and $j$ is attractive 
\cite{kk92}. Both models have the total momentum 
$P_T = \sum_{j=1}^N p^{j}_t$ as a conservative quantity
and they were extensively studied in \cite{kk92,kk94}, which 
characterized the system by the existence of an ordering process 
called cluster, which will be discussed later.

\section{Stickiness and the FTLEs distributions}
\label{method}

\begin{figure*}[htb]
  \centering
  \includegraphics*[width=0.98\columnwidth]{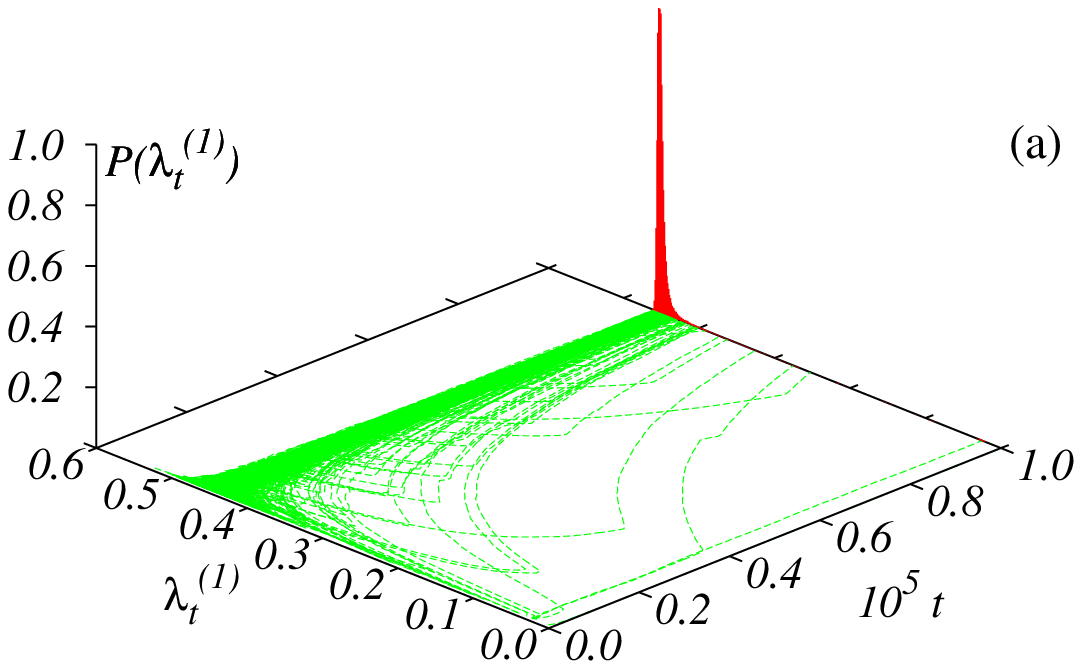}
  \includegraphics*[width=0.98\columnwidth]{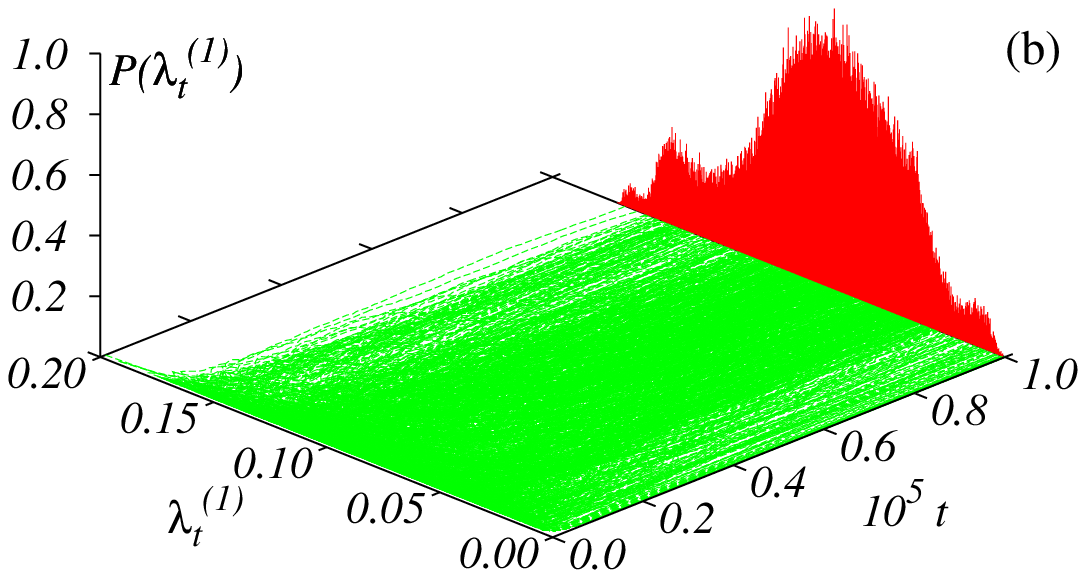}
  \caption{(Color online) Local FTLEs as a function of time for 
    $N=4$, showing their dependence on $10^3$ distinct ICs. 
    After the last iteration the FTLEs distribution over the ICs
    is plotted in red (for better visualization we used $10^6$
    ICs.)  In this example we used global coupling with (a) 
    $K=1.0$ and (b) $K=0.21$. }
  \label{N4-FTLE}
\end{figure*}

In this Section we summarize the method used to detect
stickiness from the FTLEs distribution.
Sticky motion reduces the local FTLE along a chaotic trajectory.
This can be seen in Fig.~\ref{N4-FTLE}, where the FTLEs 
are plotted as a function of time. In this case simulations were 
realized for the GC with $N=4$, $K=1.0$ [Fig.~\ref{N4-FTLE}(a)] 
and $K=0.21$ [Fig.~\ref{N4-FTLE}(b)]. 
After the last iteration the corresponding FTLEs distribution 
over initial conditions is plotted (in red). As time 
increases the local FTLE increases for a chaotic trajectory, 
but since it may touch or penetrate regular structures in the phase
space, sticky motion 
occurs and the local FTLE decreases (see green decreasing lines). 
When this trajectory leaves the sticky region, the local FTLE 
starts to increase again. Thus, those chaotic trajectories 
affected by the sticky motion will have smaller values of 
their FTLEs at the last iteration. This induces a small tail 
to the right in the distribution of the FTLEs 
[see Fig.~\ref{N4-FTLE}(a)], and it becomes asymmetric. Therefore, 
while for intermediate values of $K$ 
such asymmetric fat tails are expected, for larger $K$ values, 
where a totally chaotic motion occurs, a Gaussian distribution 
for the FTLEs is expected. 
For small values of $K$ a large amount of regular motion
in phase space is expected and thus strong sticky motion. 
This can be observed, for example, in Fig.~\ref{N4-FTLE}(b) 
for $K=0.21$. In such cases the FTLEs distribution is not 
Gaussian-like and it can be multimodal. {  This is 
a very complicated region to be characterized in general 
since only a tiny portion of ICs lead to chaotic motion, and
these are responsible for the Arnold diffusion and produce
Arnold stripes in the {\it mixed} plots ICs {\it versus} nonlinear 
parameter \cite{marcelo11-1,marcelo12-1}}.

For the cases where the FTLE distribution is a Gaussian-like 
function, higher cummulants from this distribution 
are very efficient to detect tiny sticky motion, as was 
shown for the standard map \cite{steven07} and for
high-dimensional conservative systems \cite{cesar-rost11}. 
Here we show results determining the skewness, 
or the asymmetry of the FTLEs distribution. The 
{\it skewness} is defined by $\kappa_3=
\displaystyle{\left<(\lambda_t-\left<\lambda_t\right>)^3
\right>}/{\tilde\sigma^{3/2}}$, where  
$\sigma=\tilde\sigma/\left<\lambda_t\right>^2$ 
is the relative variance and $\lambda_t$ is the local FTLE. 
For  $\kappa_3=0$ we have the regular Gaussian distribution, 
expected for a chaotic system. Since sticky motion usually 
reduces the FTLEs the asymmetry of the distribution leads to 
$\kappa_3<0$.  Thus for any  $\kappa_3<0$ sticky motion is 
expected. The flatness (forth cummulant) could also be used 
instead the skewness, but results are similar. For more details 
we refer the readers to \cite{cesar-rost11}, where it was 
shown that the variance of the FTLEs distribution is not that 
efficient to detect sticky motion as the flatness
and skewness. 

\section{Characterizing weak chaos}
\label{results}
The transition from the integrable case ($K=0$) to the weakly 
chaotic one depends on the kind of coupling between sites. As
mentioned in the introduction, in the weakly chaotic region everything 
is complicated \cite{zas88}: the positions of the regular structures, 
their size distribution, the shapes of the structures boundaries and 
the penetration inside them due to higher dimensions. As demonstrated 
above, the method of higher cummulants of the FTLEs distribution is not 
appropriate for too small $K$ values, where the amount of regular motion 
in phase space is much larger than the chaotic one. In order to understand 
better the integrable to weakly chaotic transition we analyze FTLEs
in {\it mixed} plots as an additional tool. Such plots are very useful 
to study bifurcation diagrams in conservative systems \cite{beims13}.

\subsection{The global coupling}

\begin{figure}[htb]
  \centering
  \includegraphics*[width=1.0\linewidth]{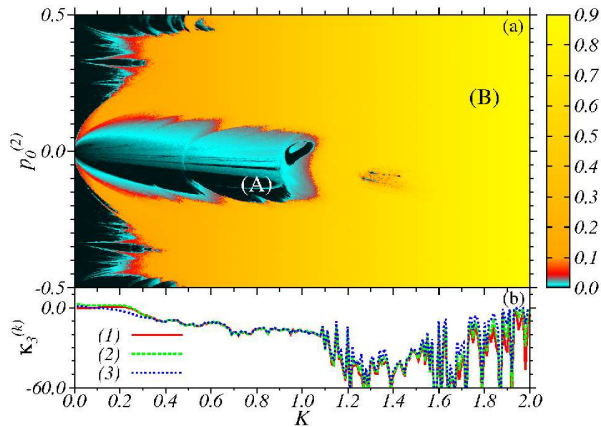}
  \caption{(Color online) Comparison of the (a) mixed plot $K\times p_0^{(2)}$
    with the (b) skewness $\kappa_3^{(k)}$ for $N=4$ and the GC
    coupling (\ref{gc}). Continuous line for $k=1$, dashed for $k=2$
    and dotted line for $k=3$.}  
\label{N4-glob}
\end{figure}

We start considering the case $N=4$ for which $8$ 
symmetric FTLEs exist, two of them are exactly zero 
due to the conservation of the total momentum. 
An interesting way to describe the complicated global dynamics 
dependence on initial conditions, and thus the sticky effect, 
is shown in the {\it mixed} plot $K\times p_0^{(2)}$ presented
in Fig.~\ref{N4-glob}(a). Other dynamical quantities ($p_0^{(i)}$, 
with $i=1,2,3,\ldots$ could also be used instead $p_0^{(2)}$). 
Colors are the largest FTLEs after $10^4$ iterations. Initial 
conditions are chosen on the invariant structure $P_T=0.0$ and 
$X_{CM}=\sum_{j=1}^NX_0^{(j)}=0.0$.
This plot shows, as a function of $K$, which 
initial conditions generate zero or positive FTLEs.
A rich variety of structures is observed. Essentially two distinct larger
regions of motions are observed and demarked in the plot as: (A) where 
FTLEs are smaller and (B) where FTLEs are larger. In between the FTLEs mix 
themselves along complicated and apparently fractal structures. For $K\to0$ 
FTLEs go to zero inside region (A). However there are some stripes, 
which emanate from $K\sim 0$ for which FTLEs are larger. Two larger 
stripes are born close to $p_0^{(0)}\sim 0.0$ and growth symmetrically 
around this point as $K$ increases. Such stripes are intervals of initial 
conditions inside which Arnold diffusion occurs and are thus named as
{\it Arnold} or {\it Chaotic} stripes. They exist for many intervals of 
initial conditions, as described in more details in 
\cite{marcelo11-1,marcelo12-1}, where their apparency was also 
observed in open billiards with rounded  corner/borders.

Along the line $p_0^{(2)}\sim 0.0$ 
the dynamics is almost regular for $K<1$, becoming chaotic for larger 
values of $K$. This picture shows us clearly that for the same $K$ value, 
different regions of the phase space (in this case $p_0^{(2)}$) have 
distinct FTLEs. Smaller FTLEs are related to those initial conditions 
which started exactly on a regular trajectory, or on a chaotic trajectory 
which touched for a finite time the regular structures from the 
high-dimensional phase space. Such regular structures can be 
global (as the invariant $P_T$), local, or collective ordered 
states which live in the high-dimensional phase space. As trajectories 
itinerate between ordered and random states, the regular structures 
affect locally the FTLEs inducing sticky motion, so that the 
corresponding FTLE decreases. Thus each point in the mixed plot from 
Fig.~\ref{N4-glob}(a) which has smaller FTLEs for a fixed $K$ value, 
is necessarily related to sticky motion. 

This is confirmed by comparing Fig.~\ref{N4-glob}(a) and (b), where
it is possible to associate stickiness $\kappa_3^{(k)}<0.0$ in 
Fig.~\ref{N4-glob}(b) 
with the existence of initial conditions which generate close to zero
FTLEs in  Fig.~\ref{N4-glob}(a). 
Figure \ref{N4-glob}(b) shows the skewness related 
to the distribution for the three remaining positive 
FTLEs as a function of $K$ for the GC. In this case 
$10^4$ random equally distributed initial conditions 
are used and $10^6$ iterations. For very small values 
of $K\lesssim 0.2$, we checked that the FTLEs 
distributions are not Gaussian-like, so that $\kappa_3^{(k)}$ 
is not well defined. For $K>0.2$ it can be seen in 
Fig.~\ref{N4-glob}(b) that $\kappa_3^{(k)}\lesssim 0.0$ for 
all three FTLEs. Thus all FTLEs  
are equally affected by the regular structures leaving 
to sticky motion. Small changes between distinct 
$\kappa_3^{(k)}$ are observed close to $K\sim 2.0$ but since 
$\kappa_3^{(k)}\to 0.0$, no relevant sticky motion is expected 
to occur anymore and the dynamics can be characterized 
as totally chaotic for $K\gtrsim 2.0$.

We can identify a clear transition 
between three dynamical regions: strongly {\it regular} for $K\lesssim 0.3$, 
where $\kappa_3^{(k)}$ is probably not well defined and many initial 
conditions generate small FTLEs; strongly {\it mixed} for 
$0.3\lesssim K \lesssim 1.8$ where many sticky regions are
observed $\kappa_3^{(k)}<0.0$ and; the {\it chaotic} one for $K\gtrsim 1.8$ 
where $\kappa_3^{(k)}\to 0.0$. Interesting to mention is that in the 
mixed region we observe the common motion, where all FTLEs are
equally affected by the regular structures in the high-dimensional
phase space.

\begin{figure}[htb]
  \centering
  \includegraphics*[width=1.0\linewidth]{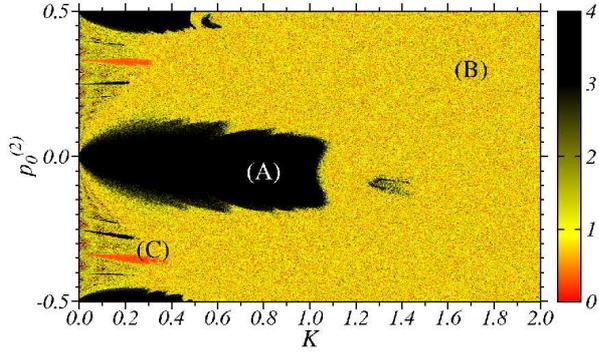}
  \caption{(Color online) Order parameter (see colors) shown for the 
{\it mixed} plot $K\times p_0^{(2)}$ for $N=4, t=10^4$, for the GC.}
  \label{N4-glob-order}
\end{figure}

\begin{figure}[htb]
  \centering
  \includegraphics*[width=1.0\linewidth]{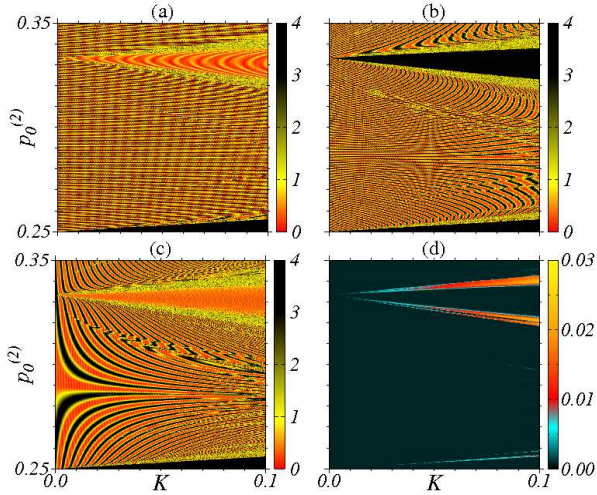}
  \caption{(Color online) Magnification from Fig.~\ref{N4-glob-order} (see 
white box) for four different times: (a) for $t=1.0\times10^3$ (b) for 
$t=9.0\times10^3$, (c) for $t=1.0\times10^4$ and (d) FTLE.}
  \label{N4-glob-order-zoom}
\end{figure}

To compare our results with  the ordering states shown in 
\cite{kk94}, we firstly write down the order parameter

\begin{equation}
Z_t=\frac{1}{\sqrt{N}}\sum_{j=1}^N exp{\left[2\pi i x_t^{(j)}\right]}.
\label{Z}
\end{equation}
From this parameter it is possible to distinguish essentially three 
motions \cite{kk94}:
(i) when $Z_t=N$, {\it i.e.} all $x_t^{(j)}$ are the same; (ii) when $Z_t\approx 1$, 
{\it i.e.~}the $x_t^{(j)}$ are randomly distributed and (iii) when $Z_t=0$  which
occurs if $x_t^{(j)}$ are evenly spaced. States (i) and (iii) are denominated
ordered states while (ii) a chaotic (or random) state. It is essential to 
recognize that $Z_t$ is a {\it local} quantity and {\it changes} in time, 
oscillating between ordered and random states. In this sense, it is not so 
powerful to make general statements about the dynamics as the FTLEs 
distribution. Independent of that, it is interesting to look at the 
order parameter in the {\it mixed} plot $K\times p_0^{(2)}$, which gives 
us a clear picture of the dynamics dependence on the ICs. This 
is shown in Fig.~\ref{N4-glob-order} where the colors represent
the order parameter. ICs are along the conservative quantity 
$P_T=0$ with center of mass $X_{CM}=0$. It is possible to distinguish 
three main regions: (A) where the order parameter $Z_t$ is close do $N=4$ (see 
color bar) meaning that a fully clustered motion is expected; (B) where 
$Z_t\sim 1.0$ and a random state is expected and (C) where $Z_t\sim 0.0$ and 
an ordered state is expected. The {\it mixed} plot very clearly relates the 
initial momentum in the phase space and the corresponding dynamics. 
Although the dynamics is well understood by such {\it mixed} plot, 
the order parameter has not a definitive value since this is a high-dimensional 
system and ordered and random states can alternate in time. To show the 
complexity of the dynamics in time, Fig.~\ref{N4-glob-order-zoom} shows a 
magnification of Fig.~\ref{N4-glob-order} for three different exemplary times: 
(a)  $t=1.0\times10^3$ (b) $t=9.0\times10^3$, (c) $t=1.0\times10^4$ 
and (d) FTLE for  $t=1.0\times10^4$.
Basically we identify three regions in Fig.~\ref{N4-glob-order-zoom}(a): 
the large region with alternating values of $Z_t$ and two stripes, one
red ($Z_t\sim 0.0$) and one black ($Z_t\sim 4.0$). These stripes increase as $K$
increases. However, as times goes on ($t=9.0\times10^3$) 
Fig.~\ref{N4-glob-order-zoom}(b) shows that inside the upper stripe the
dynamics changes to a fully clustered state, while inside the lower stripe 
remains unchanged. As time increases to $t=1.1\times10^4$ the dynamics
inside the upper stripe changes again to $Z_t\sim 0.0$, showing the 
alternating dynamics between the two ordered states connected in time
by the random state. The large region between the two stripes has now
well defined structures which alternate in space between the two ordered
states (black and red) connected by the random state (yellow), since we
always observed this color between the ordered state. Figure 
\ref{N4-glob-order-zoom}(d) shows the corresponding FTLE which will be
described below.

\begin{figure}[htb]
  \centering
 \includegraphics*[width=1.0\linewidth]{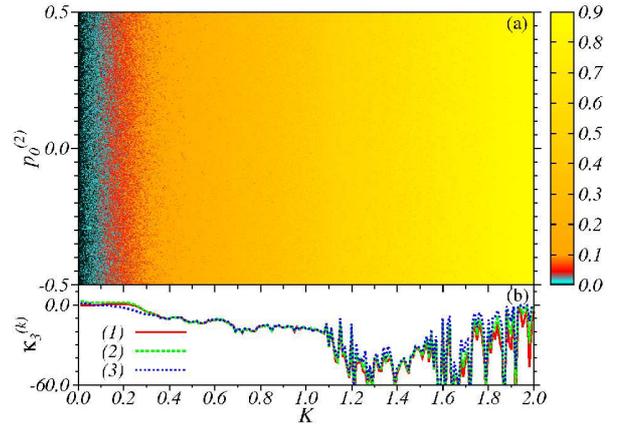}
 \caption{(Color online) The case of random ICs for (a)  the {\it mixed} 
plot $K\times p_0^{(2)}$, (b) $\kappa_3^{k}$ and (c) the order parameter.
Continuous line for $k=1$, dashed for $k=2$ and dotted line for $k=3$.}
  \label{N4-glob-ran}
\end{figure}
Comparing Figs.~\ref{N4-glob}(a) and \ref{N4-glob-order} it is possible 
to identify regions (A) and (B) showing an association between 
ordered state with small FTLEs and random states with larger FTLEs. This
association however is not definitive since the order parameter, which 
defines the ordered and random states, may change in time.
The distinction between the ordered state from regions (A) and (C) from 
Fig.~\ref{N4-glob-order} are not observed in Fig.~\ref{N4-glob}(a). 
In fact, for smaller values of $K$ the dynamics is very rich and 
complicated. This is better observed in the magnifications shown
in Fig.~\ref{N4-glob-order-zoom}. While Figs.~\ref{N4-glob-order-zoom} 
(a)-(c) show the alternating dynamics of ordered states 
($Z_t\sim0.0\to 4.0\to 0.0\to\ldots$) inside the stripes mentioned above, 
Fig.~\ref{N4-glob-order-zoom}(d) shows the FTLEs. Inside these stripes the 
FTLEs increase, meaning 
that the ordered states ($Z_t\sim 0.0,4.0$) present chaotic dynamics 
because they are connected through the random states which have 
($Z_t\sim 1.0$). Outside the stripes the dynamics is very close to 
non-chaotic.

Although all figures and quantities discussed above nicely explain 
the dynamics, important changes occur when initial conditions are varied.
For example, for the above discussed cases we always start initial
conditions along the constant invariant structure $P_T=0$ and $X_{CM}=0$. 
By increasing the value of $P_T$ we observed (not shown) that region ($B$) 
and ($C$) from Figs.~\ref{N4-glob-order} and \ref{N4-glob} start to decrease. 
This clearly means that when  $P_T$ increases, the sticky time for 
which trajectories are trapped to invariant structures in phase 
space decreases. Moreless, the dynamics in the {\it mixed} plots changes 
drastically when initial conditions are chosen {\it away} from 
the invariant $P_T$. This is shown in Fig.~\ref{N4-glob-ran}(a) for the FTLEs.
Initial conditions are chosen randomly. Very fast we 
recognize that almost only region (B) remains in the {\it mixed} plot. 
This agrees with the statements mentioned in \cite{kk94}, that as initial 
randomness increases the duration of ordered state decreases rapidly.
On the other hand, the FTLEs shown in Fig.~\ref{N4-glob-ran}(a) already 
show that for small $K$ values the FTLEs decrease. But no stripes 
or well defined region with specific dynamics are observed. Clearly as
$K$ increases, the amount of points related to sticky motion decreases
and close to $K\sim 2.0$ only yellow points are observed, meaning that
a strong chaotic behavior is expected. This is very easily confirmed 
by the $\kappa_3^{(k)}$, which tends to zero for $K\sim 2.0$. In distinction
to the plot Fig.~\ref{N4-glob-ran}(a) the stickiness effect is 
accurately detected by $\kappa_3^{(k)}$ in Fig.~\ref{N4-glob-ran}(b).

\begin{figure}[htb]
  \centering
\includegraphics*[width=1.0\linewidth]{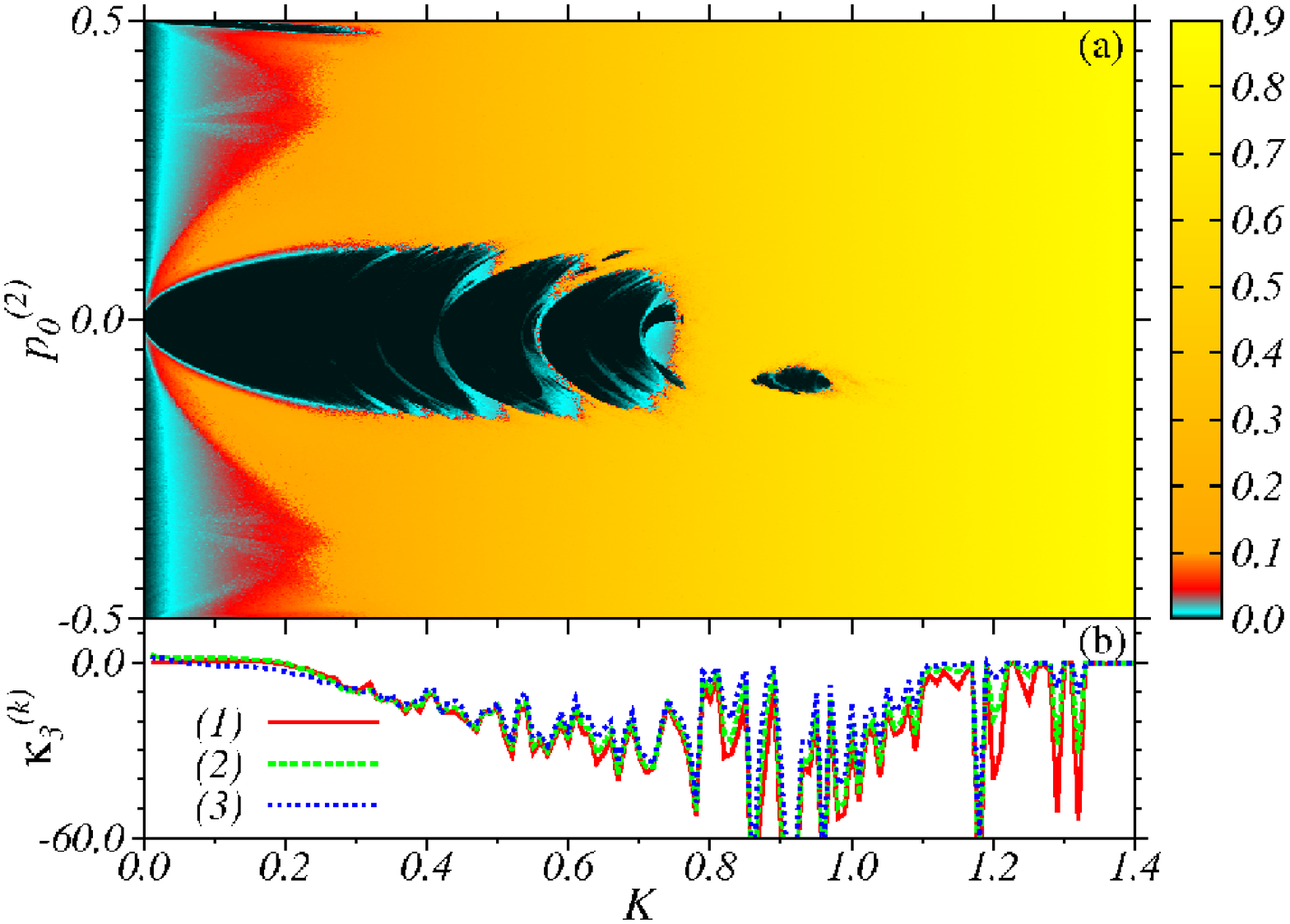}
 \caption{(Color online) Comparation of the (a) {\it mixed} plot 
$K\times p_0^{(2)}$ with the (b) skewness $\kappa_3^{(k)}$ for $N=4$ and the 
LC coupling (\ref{lc}). Continuous line for $k=1$, dashed for $k=2$ and 
dotted line for $k=3$.} 
  \label{N4-loc}
\end{figure}

Extensive numerical simulations were realized for values of $N=16$ 
and the main relevant observation is that for reasonable values
of the nonlinear parameter $K\sim 0.10$, the dynamics is almost chaotic. 
For smaller values o $K\lesssim 0.03$ the regular portion of the phase 
space is so large that non-Gaussian FTLEs were found, making our 
method not appropriate for the analysis.

\subsection{Local coupling}

Now we continue to discuss the case $N=4$ but for the LC from 
Eq.~(\ref{lc}). Figure \ref{N4-loc}(b) shows the skewness 
related to the distribution for the three  positive FTLEs as 
a function of $K$. As for the GC numerical investigations, 
$10^4$ random equally distributed initial conditions 
are used and $10^6$ iterations. For values of $K\lesssim 0.2$ 
the FTLEs distributions are not Gaussian-like, so that $\kappa_3^{(k)}$ 
is not well defined. For $0.2\lesssim K\lesssim1.1$ it can be 
seen in Fig.~\ref{N4-loc}(b) 
that $\kappa_3^{(k)}<0.0$ for all three FTLEs. The FTLEs for 
the three unstable directions are almost equally affected 
by the regular structures leaving to sticky common motion. 
Tiny changes between distinct $\kappa_3^{(k)}$ are observed. 
For $K>1.1$,  $\kappa_3^{(k)}\to 0.0$ but close to $K\sim 1.2,1.3$
we see effects of sticky motion. Figure \ref{N4-loc}(a)
presents the mixed plot $K\times p_0^{(2)}$, where colors 
are the largest FTLEs after $10^5$ iterations. As in 
Fig.~\ref{N4-glob}(a), initial conditions are chosen on the 
invariant structure $P_T=0.0$ and $X_{CM}=0.0$. 
Comparing with Fig.~\ref{N4-glob}(a) we observe that the chaotic region
and the sticky region close to $ p_0^{(2)}$ remain almost unaltered. The
essential differences occur for very small values of $K\lesssim 0.3$, where
FTLEs are a small amount larger. Since we consider here the LC, the small 
perturbation in $K$ is not able to destroy the regular dynamics from the 
GC. In addition, the chaotic stripes observed Fig.~\ref{N4-glob}(a) become 
blurred and the Arnold web is expected to be easier destroyed. By 
comparing Fig.~\ref{N4-loc}(a) and (b), where again
we can easily associate stickiness $\kappa_3^{(k)}<0.0$ in Fig.~\ref{N4-loc}(b) 
with the existence of initial conditions which generate close to zero
FTLEs in  Fig.~\ref{N4-loc}(a). The transition from strongly regular for 
$K\lesssim 0.2$, mixed for $0.3\lesssim K \lesssim 1.1$ and chaotic one 
for $K\gtrsim 1.1$ can be identified. For the particular values 
$K\sim 1.17,1.2,1.29,1.31$ the asymmetry $\kappa_3^{(k)}$ is negative in 
Fig.~\ref{N4-loc}(b) and sticky motion should be present. Such points cannot
be identified in Fig.~\ref{N4-loc}(a), showing the power of the FTLEs 
distribution to detect the sticky motion.

For the case of random chosen ICs the chaotic stripes disappear
and results are similar to those of Fig.\ref{N4-glob-ran}. In other
words, the choice of ICs is also essential for the LC.
Also for the LC we studied the case $N=16$ and 
results are almost similar to those from the GC. 

\section{Conclusions}
\label{conclusion}

The full characterization of the dynamics in high-dimensional Hamiltonian
systems is a very difficult task. The restriction of plots to $2-$ and 
$3-$dimensions, the FTLEs dependence on initial conditions and sticky 
effects are the greater obstacles to solve this task for weakly chaotic 
systems.  While the order parameter, studied extensively in \cite{kk92,kk94} 
for the coupled maps analyzed here, contains time dependent informations 
about dynamical states, the suitable quantities to fully characterize the 
dynamics are the higher order cummulants of the FTLEs distribution. In 
this work we only discuss results using the third cummulant, the skewness. 
Identical statements follow if one uses the forth cummulant, the kurtosis.  
The skewness gathers informations about  ordered and random states and 
stickiness for each (un)stable directions from the phase space. In 
distinction to a previous work \cite{cesar-rost11}, here the characterization 
is performed to local and global couplings of $N=4$ maps in a conservative 
Hamiltonian system. Comparison with the order parameter were also realized.
For very small perturbations the FTLEs distributions are not Gaussian-like and 
the skewness cannot be defined.  For intermediate values of the perturbations
the mixed dynamics was characterized for both, local and global couplings. 
It was observed that the common motion (i.e., that regular structures in 
phase space ``attract'' the chaotic trajectories by the same amount in all 
unstable directions) occurs independently of the coupling 
length. Additional numerical simulations have shown that the common motion 
tends to dissapear for a larger number of coupled maps and larger nonlinear 
parameter values. It was also shown that FTLEs shown in mixed plots (initial 
condition of the dynamical variable {\it versus} nonlinear parameter) are very 
convenient pictures to study the initial conditions dependence of the dynamics.
Initial conditions which start inside invariant structures (total momentum 
$P_T=0$, in our case) tend to be regular for very long times, and sticky 
effects are of most importance. On the other hand, when initial conditions 
are chosen randomly, stickiness effects become less visible. These results 
also show the fundamental role of initial conditions in weakly chaotic systems.

\section*{Acknowledgments}
The authors thank FINEP (under project CTINFRA-1) , CM and MWB thank CNPq for 
financial support. 


\section*{References}

\begin{thebibliography}{99}
\expandafter\ifx\csname natexlab\endcsname\relax\def\natexlab#1{#1}\fi
\expandafter\ifx\csname bibnamefont\endcsname\relax
  \def\bibnamefont#1{#1}\fi
\expandafter\ifx\csname bibfnamefont\endcsname\relax
  \def\bibfnamefont#1{#1}\fi
\expandafter\ifx\csname citenamefont\endcsname\relax
  \def\citenamefont#1{#1}\fi
\expandafter\ifx\csname url\endcsname\relax
  \def\url#1{\texttt{#1}}\fi
\expandafter\ifx\csname urlprefix\endcsname\relax\def\urlprefix{URL }\fi
\providecommand{\bibinfo}[2]{#2}
\providecommand{\eprint}[2][]{\url{#2}}

\bibitem[{\citenamefont{Chernikov et~al.}(1988)\citenamefont{Chernikov,
  Sagdeev, and Zaslavsky}}]{zas88}
\bibinfo{author}{\bibfnamefont{A.A.} \bibnamefont{Chernikov}},
  \bibinfo{author}{\bibfnamefont{R.Z.} \bibnamefont{Sagdeev}},
  \bibnamefont{and} \bibinfo{author}{\bibfnamefont{G.M.}
  \bibnamefont{Zaslavsky}}, \bibinfo{journal}{Physics Today}
  {\bibinfo{volume}{41}} (\bibinfo{year}{1988}) \bibinfo{pages}{27}.

\bibitem[{\citenamefont{Meiss}(1992)}]{meiss92}
\bibinfo{author}{\bibfnamefont{J.D.} \bibnamefont{Meiss}},
  \bibinfo{journal}{Rev.~Mod.~Phys.} {\bibinfo{volume}{64}} 
 (\bibinfo{year}{1992}) \bibinfo{pages}{795}.

\bibitem[{\citenamefont{Richter et~al.}(arXiv:1307.6109)\citenamefont{Richter,
  Lange, B\"acker, and Ketzmerick}}]{martin13}
\bibinfo{author}{\bibfnamefont{M.}~\bibnamefont{Richter}},
  \bibinfo{author}{\bibfnamefont{S.}~\bibnamefont{Lange}},
  \bibinfo{author}{\bibfnamefont{A.}~\bibnamefont{B\"acker}}, \bibnamefont{and}
  \bibinfo{author}{\bibfnamefont{R.}~\bibnamefont{Ketzmerick}}
  (\bibinfo{year}{arXiv:1307.6109}).

\bibitem[{\citenamefont{Zaslavski}(2002)}]{zas02}
\bibinfo{author}{\bibfnamefont{G.M.} \bibnamefont{Zaslavski}},
  \bibinfo{journal}{Phys.~Rep.} {\bibinfo{volume}{371}} (\bibinfo{year}{2002})
  \bibinfo{pages}{461}.

\bibitem[{\citenamefont{Livorati et~al.}(2012)}]{denis12}
\bibinfo{author}{\bibfnamefont{A.L.P.} \bibnamefont{Livorati}},
\bibinfo{author}{\bibfnamefont{T.} \bibnamefont{Kroetz}},
\bibinfo{author}{\bibfnamefont{C.P.} \bibnamefont{Dettmann}},
\bibinfo{author}{\bibfnamefont{I.L.} \bibnamefont{Caldas}},
\bibinfo{author}{\bibfnamefont{E.D.} \bibnamefont{Leonel}},
  \bibinfo{journal}{Phys.~Rev.~E} {\bibinfo{volume}{86}} (\bibinfo{year}{2012})
  \bibinfo{pages}{036203}.

\bibitem[{\citenamefont{Pettini and Vulpiani}(1984)}]{pettini84}
\bibinfo{author}{\bibfnamefont{M.}~\bibnamefont{Pettini}} \bibnamefont{and}
  \bibinfo{author}{\bibfnamefont{A.}~\bibnamefont{Vulpiani}},
  \bibinfo{journal}{Phys.~Lett.} {\bibinfo{volume}{106A}}  (\bibinfo{year}{1984})
  \bibinfo{pages}{207}.

\bibitem[{\citenamefont{K.Tsiganis et~al.}(2000)\citenamefont{K.Tsiganis,
  A.Anastasiadis, and H.Varvoglis}}]{tsiganis00}
\bibinfo{author}{\bibnamefont{K. Tsiganis}},
  \bibinfo{author}{\bibnamefont{A. Anastasiadis}}, \bibnamefont{and}
  \bibinfo{author}{\bibnamefont{H. Varvoglis}}, \bibinfo{journal}{Chaos,
  Solitons and Fractals} {\bibinfo{volume}{11}}  (\bibinfo{year}{2000}) \bibinfo{pages}{2281}.

\bibitem[{\citenamefont{Malagoli et~al.}(1986)\citenamefont{Malagoli, Paladin,
  and Vulpiani}}]{malagoli86}
\bibinfo{author}{\bibfnamefont{A.}~\bibnamefont{Malagoli}},
  \bibinfo{author}{\bibfnamefont{G.}~\bibnamefont{Paladin}}, \bibnamefont{and}
  \bibinfo{author}{\bibfnamefont{A.}~\bibnamefont{Vulpiani}},
  \bibinfo{journal}{Phys.~Rev.~A} {\bibinfo{volume}{34}} (\bibinfo{year}{1986})
  \bibinfo{pages}{1550}.

\bibitem[{\citenamefont{Froyland and Padberg}(2009)}]{froyland09}
\bibinfo{author}{\bibfnamefont{G.}~\bibnamefont{Froyland}} \bibnamefont{and}
  \bibinfo{author}{\bibfnamefont{K.}~\bibnamefont{Padberg}},
  \bibinfo{journal}{Physica D} {\bibinfo{volume}{238}}  (\bibinfo{year}{2009})
  \bibinfo{pages}{1507}.

\bibitem[{\citenamefont{Dellnitz and Junge}(1997)}]{dellnitz97}
\bibinfo{author}{\bibfnamefont{M.}~\bibnamefont{Dellnitz}} \bibnamefont{and}
  \bibinfo{author}{\bibfnamefont{O.}~\bibnamefont{Junge}},
  \bibinfo{journal}{Int.~J.~Bif.~and Chaos} {\bibinfo{volume}{7}}
 (\bibinfo{year}{1997}) \bibinfo{pages}{2475}.

\bibitem[{\citenamefont{Skokos et~al.}(2004)\citenamefont{Skokos, Antonopoulos,
  Bountis, and M.N.Vrahatis}}]{skokos04}
\bibinfo{author}{\bibfnamefont{C.}~\bibnamefont{Skokos}},
  \bibinfo{author}{\bibfnamefont{C.}~\bibnamefont{Antonopoulos}},
  \bibinfo{author}{\bibfnamefont{T.C.} \bibnamefont{Bountis}},
  \bibnamefont{and} \bibinfo{author}{\bibnamefont{M.N. Vrahatis}},
  \bibinfo{journal}{J.~Phys.~A} {\bibinfo{volume}{37}}  (\bibinfo{year}{2004})
  \bibinfo{pages}{6269}.

\bibitem[{\citenamefont{Antonopoulos et~al.}(2006)\citenamefont{Antonopoulos,
  Bountis, and Skokos}}]{skokos06}
\bibinfo{author}{\bibfnamefont{C.}~\bibnamefont{Antonopoulos}},
  \bibinfo{author}{\bibfnamefont{T.}~\bibnamefont{Bountis}}, \bibnamefont{and}
  \bibinfo{author}{\bibfnamefont{C.}~\bibnamefont{Skokos}},
  \bibinfo{journal}{Int.~J.~Bif.~and Chaos} {\bibinfo{volume}{16}} (\bibinfo{year}{2004})
  \bibinfo{pages}{1777}.

\bibitem[{\citenamefont{Tomsovic and Lakshminarayan}(2007)}]{steven07}
\bibinfo{author}{\bibfnamefont{S.}~\bibnamefont{Tomsovic}} \bibnamefont{and}
  \bibinfo{author}{\bibfnamefont{A.}~\bibnamefont{Lakshminarayan}},
  \bibinfo{journal}{Phys.~Rev.~E} {\bibinfo{volume}{76}}  (\bibinfo{year}{2007})
  \bibinfo{pages}{036207}.

\bibitem[{\citenamefont{C.Manchein et~al.}(2012)\citenamefont{C.Manchein,
  M.W. Beims, and J.M. Rost}}]{cesar-rost11}
\bibinfo{author}{\bibnamefont{C. Manchein}},
  \bibinfo{author}{\bibnamefont{M.W. Beims}}, \bibnamefont{and}
  \bibinfo{author}{\bibnamefont{J.M. Rost}}, \bibinfo{journal}{Chaos}
  {\bibinfo{volume}{22}}  (\bibinfo{year}{2012}) \bibinfo{pages}{033137}.

\bibitem[{\citenamefont{Woellner et~al.}(2009)\citenamefont{Woellner, Lopes,
  Viana, and Caldas}}]{woellner09}
\bibinfo{author}{\bibfnamefont{C.F.} \bibnamefont{Woellner}},
  \bibinfo{author}{\bibfnamefont{S.R.} \bibnamefont{Lopes}},
  \bibinfo{author}{\bibfnamefont{R.L.} \bibnamefont{Viana}}, \bibnamefont{and}
  \bibinfo{author}{\bibfnamefont{I.L.} \bibnamefont{Caldas}},
  \bibinfo{journal}{Chaos, Solitons and Fractals}
  {\bibinfo{volume}{41}} (\bibinfo{year}{2009}) \bibinfo{pages}{2201}.

\bibitem[{\citenamefont{K.Kaneko and T.Konishi}(1989)}]{kk89}
\bibinfo{author}{\bibnamefont{K. Kaneko}} \bibnamefont{and}
  \bibinfo{author}{\bibnamefont{T. Konishi}}, \bibinfo{journal}{Phys. Rev. A}
  {\bibinfo{volume}{40}}  (\bibinfo{year}{1989}) \bibinfo{pages}{6130}.

\bibitem[{\citenamefont{T.Konishi and K.Kaneko}(1990)}]{kk90}
\bibinfo{author}{\bibnamefont{T. Konishi}} \bibnamefont{and}
  \bibinfo{author}{\bibnamefont{K. Kaneko}}, \bibinfo{journal}{J. Phys. A}
  {\bibinfo{volume}{23}}  (\bibinfo{year}{1990}) \bibinfo{pages}{715}.

\bibitem[{\citenamefont{M.S.Custodio et~al.}(2012)\citenamefont{M.S.Custodio,
  C.Manchein, and M.W.Beims}}]{marcelo12-1}
\bibinfo{author}{\bibnamefont{M.S. Custodio}},
  \bibinfo{author}{\bibnamefont{C. Manchein}}, \bibnamefont{and}
  \bibinfo{author}{\bibnamefont{M.W. Beims}}, \bibinfo{journal}{Chaos}
  {\bibinfo{volume}{22}}   (\bibinfo{year}{2012}) \bibinfo{pages}{026112}.

\bibitem[{\citenamefont{T.Konishi and K.Kaneko}(1992)}]{kk92}
\bibinfo{author}{\bibnamefont{T. Konishi}} \bibnamefont{and}
  \bibinfo{author}{\bibnamefont{K. Kaneko}}, \bibinfo{journal}{J. Phys. A}
  {\bibinfo{volume}{1992}}  (\bibinfo{year}{1992}) \bibinfo{pages}{6283}
 .

\bibitem[{\citenamefont{Kaneko and Konishi}(1994)}]{kk94}
\bibinfo{author}{\bibfnamefont{K.}~\bibnamefont{Kaneko}} \bibnamefont{and}
  \bibinfo{author}{\bibfnamefont{T.}~\bibnamefont{Konishi}},
  \bibinfo{journal}{Physica~D} {\bibinfo{volume}{71}}  (\bibinfo{year}{1994})
  \bibinfo{pages}{146}.

\bibitem[{\citenamefont{Custodio and Beims}(2011)}]{marcelo11-1}
\bibinfo{author}{\bibfnamefont{M.S.} \bibnamefont{Custodio}} \bibnamefont{and}
  \bibinfo{author}{\bibfnamefont{M.W.} \bibnamefont{Beims}},
  \bibinfo{journal}{Phys.~Rev.~E} {\bibinfo{volume}{83}}  (\bibinfo{year}{2011})
  \bibinfo{pages}{056201}.

\bibitem[{\citenamefont{Manchein and Beims}(2013)}]{beims13}
\bibinfo{author}{\bibfnamefont{C.}~\bibnamefont{Manchein}} \bibnamefont{and}
  \bibinfo{author}{\bibfnamefont{M.W.} \bibnamefont{Beims}},
  \bibinfo{journal}{Phys.~Lett.~A} {\bibinfo{volume}{377}} 
 (\bibinfo{year}{2013}) \bibinfo{pages}{789}.
\end{thebibliography}

\end{document}